%% file: planckbubblevPostAandAReport2.tex
\documentclass[twocolumn,traditabstract]{aa}
\usepackage{amssymb}
\usepackage{graphicx,amsmath}
\usepackage[hyperindex,breaklinks=true, colorlinks, citecolor=blue]{hyperref}  
\usepackage[varg]{txfonts}
\usepackage{dblfloatfix}
\usepackage{bm}
\usepackage{threeparttable}
\usepackage{multirow}
\providecommand{\sorthelp}[1]{}

\newcommand{\juan}[1]{{#1}}
\newcommand{\hjuan}[1]{}

\newcommand{\handrea}[1]{}

\include{Planck}

\newcommand{\planck}{\Planck}  

\begin{document}
\title{The magnetic environment of the Orion-Eridanus superbubble as revealed by {\it Planck}}
\authorrunning{Soler, Bracco, \& Pon}
\titlerunning{Magnetic fields in the Orion-Eridanus Superbubble}

\author{J. D. Soler\inst{\ref{MPIA}}
\and A. Bracco\inst{\ref{Nordita},\ref{AIM}}
\and A. Pon\inst{\ref{UWO}}}

\institute{Max-Planck-Institute for Astronomy, K\"{o}nigstuhl 17, 69117 Heidelberg, Germany. \email{soler@mpia.de} \label{MPIA}
\and Nordita, KTH Royal Institute of Technology and Stockholm University, Roslagstullsbacken 23, 10691 Stockholm, Sweden. \\ \email{andrea.bracco@su.se}\label{Nordita}
\and Laboratoire AIM, Paris-Saclay, CEA/IRFU/SAp - CNRS - Universit\'{e} Paris Diderot, 91191, Gif-sur-Yvette Cedex, France.\label{AIM} 
\and Department of Physics and Astronomy, University of Western Ontario, 1151 Richmond Street, London, ON, Canada, N6A 3K7. \label{UWO}}

\abstract{ 
Using the 353-GHz polarization observations by the \planck\ satellite we characterize the magnetic field in the Orion-Eridanus superbubble, a nearby expanding structure that spans more than 1600 square degrees in the sky. 
We identify a region of both low dispersion of polarization orientations and high polarization fraction \juan{associated} with the outer wall of the superbubble \juan{identified in the most recent models of the large-scale shape of the region}.  
We use the Davis-Chandrasekhar-Fermi method to derive plane-of-the-sky magnetic field strengths of tens of $\mu$G toward the southern edge of the bubble.
The comparison of these values with existing Zeeman splitting observations of H{\sc i} in emission suggests that \juan{the large-scale magnetic field in the region was primarily shaped by the expanding superbubble}.
}


\date{Received 01 Nov 2017 / Accepted 08 Dec 2017}

\maketitle

\section{Introduction}\label{introduction}
The Orion-Eridanus superbubble is a large circular structure \juan{with a diameter close to} 45\deg\ \citep{Heiles76,Reynolds79,heiles1999}.
It is most likely the result of the combined effects of ionizing UV radiation, stellar winds, and a sequence of supernova (SN) explosions from the Orion OB association \citep{Burrows93,Brown95,Guo95,Bally08}.
Due to its proximity \juan{($d$\,$<$\,$500$\,pc)} and its association with the Orion molecular cloud, it has been studied in multiple wavelengths, serving as a benchmark to the study of superbubbles and of feedback in the process of star formation \citep[][and references therein]{Ochsendorf15,Pon16Ochsendorf}. 

The magnetic field of the Orion-Eridanus bubble has previously been characterized using observations of Zeeman splitting of atomic hydrogen (H{\sc i}) emission at 21-cm, which provide the strength and direction of the line-of-sight component of the magnetic field ($\vec{B}_{\parallel}$), and with optical polarization measurements, which indicate the orientation of the plane-of-the-sky component of the magnetic field \citep[$\vec{B}_{\perp}$,][]{Heiles89,heiles1997}.
Around the Orion molecular cloud, these observations show a sharp reversal of $\vec{B}_{\parallel}$, which is thought to be related to a large-scale shock sweeping over the dense clouds. 
However, a large-scale characterization of the magnetic field of the region is still missing because of the paucity of H{\sc i} Zeeman-splitting and stellar polarization observations at Galactic latitudes $b$\,$\lesssim$\,$-$22\deg.
In this letter we examine the multiphase structure of the Orion-Eridanus superbubble and \juan{its} global magnetic-field properties using, for the first time, the \planck\ all-sky polarization observations at 353\,GHz (850\,$\mu$m) \citep{planck2014-XIX}. 

\begin{figure*}
\vspace{-0.2cm}
\centerline{
\includegraphics[height=0.34\textheight]{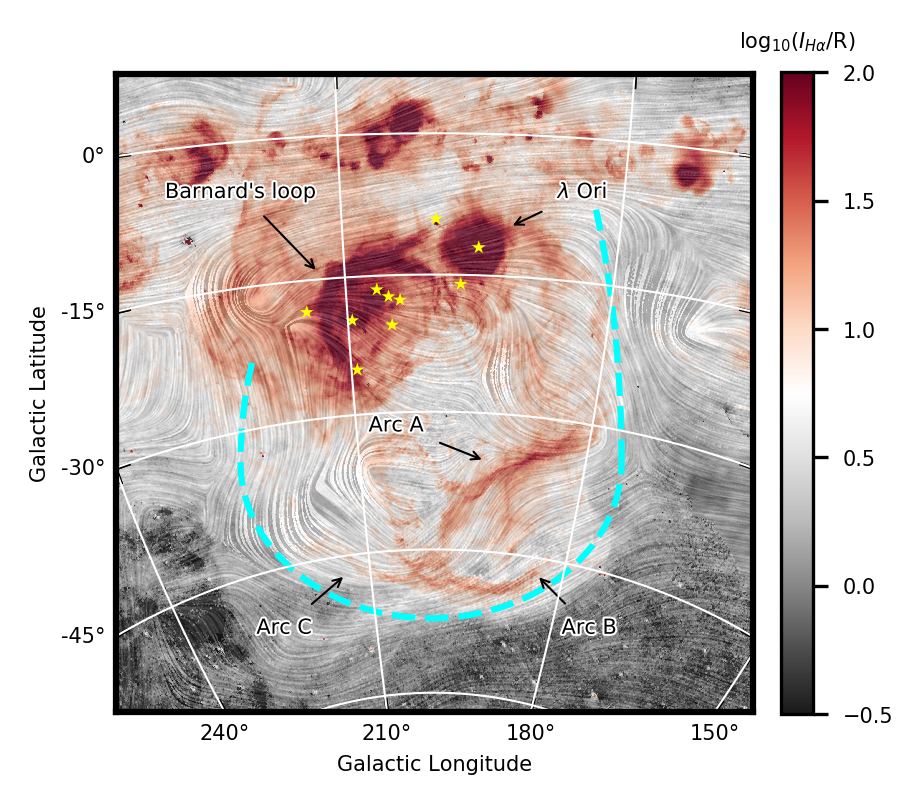}
\hspace{-0.3cm}
\includegraphics[height=0.34\textheight]{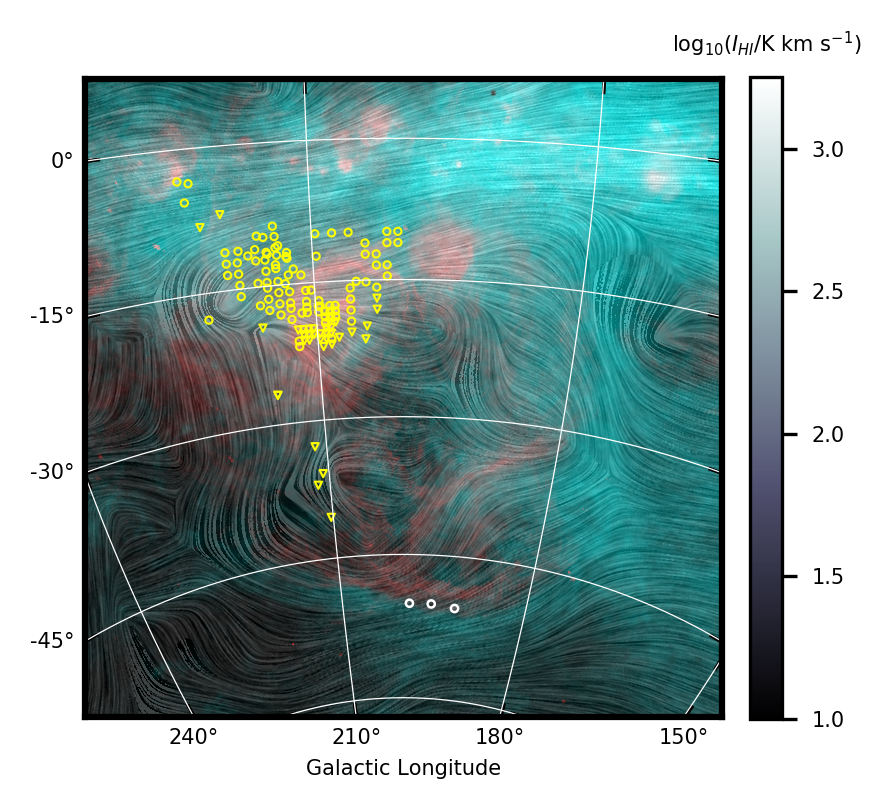}
}
\vspace{-0.25cm}
\caption{
\juan{Stereographic projections of observations toward the Orion-Eridanus superbubble.
The drapery pattern, produced using the line integral convolution technique \cite[LIC,][]{cabral1993}, corresponds to plane-of-the-sky magnetic field orientation inferred from the Planck-353\,GHz polarization observations}.
\emph{Left.} Total integrated H$\alpha$ emission map. 
The dashed line indicates the approximate location of the edge of the superbubble.
The yellow symbols correspond to the main stars in the Orion constellation.   
\emph{Right.} Total integrated H$\alpha$ emission \citep[][]{Gaustad2001} and H{\sc i} 21-cm emission \citep[][]{hi4pi2016} integrated between $-20$ and $20$\,km\,s$^{-1}$ shown in red and teal colors, respectively.
The yellow symbols correspond to the line-of-sight magnetic field directions derived from the H{\sc i} emission-line Zeeman splitting observations \citep{Heiles89,heiles1997}. 
The circles and triangles correspond to magnetic fields pointing toward and away from the observer, respectively. The three white circles in the bottom are the regions analyzed in this letter.
}\label{fig:Bfield}
\end{figure*}

\section{Superbubble morphology}\label{sec:properties}

Figure \ref{fig:Bfield} shows the orientation of $\vec{B}_{\perp}$, inferred from the \planck\ 353-GHz polarization observations \citep{planck2014-a01}, overlaid on the H$\alpha$ integrated emission from the Southern H-Alpha Sky Survey Atlas \citep[SHASSA,][]{Gaustad2001, Finkbeiner03}, and the 21-cm emission, integrated between -$20$ and $20$\,km\,s$^{-1}$ ($v_{\rm LSR}$), from the all-sky H{\sc i} survey based on the EBHIS and GASS surveys \citep[HI4PI,][]{hi4pi2016}.
The most prominent H$\alpha$ features of the superbubble, Arc~A, B, and C \citep{Johnson78}, the $\lambda$ Orionis H{\sc ii} region \citep{Mathieu2008}, and Barnard's Loop \citep{Barnard1894}, are also labelled in the right-hand-side panel of the image.

\juan{We consider} the \planck\ 353-GHz polarization observations, \juan{originally taken at 4\parcm8 resolution, smoothed with a 1\deg\ FWHM Gaussian beam} to reach \juan{a signal-to-noise ratio of at least 3 over the whole region} (see Appendix~\ref{app:Planck}).
These observations reveal three separate regions with different orientations of $\vec{B}_{\perp}$.
(i) Toward the north of the superbubble, for $b$\,$>$\,$-$15\deg, $\vec{B}_{\perp}$ appears to be relatively uniform and oriented parallel to the Galactic \juan{plane}, apparently unaffected by Barnard's loop or $\lambda$ Orionis. 
This is potentially the product of both the angular resolution of the polarization observations and the line-of-sight confusion toward the Galactic \juan{plane}. 
(ii) In the center of the superbubble, $\vec{B}_{\perp}$ appears much more \juan{tangled}. 
(iii) Along the eastern and southern edges of the superbubble, for $b$\,$<$\,$-$15\deg, $\vec{B}_{\perp}$ \juan{clearly follows} the rim of the superbubble across tens of degrees.

\begin{figure}[ht!]
\vspace{-0.2cm}
\centerline{\includegraphics[height=0.3\textheight]{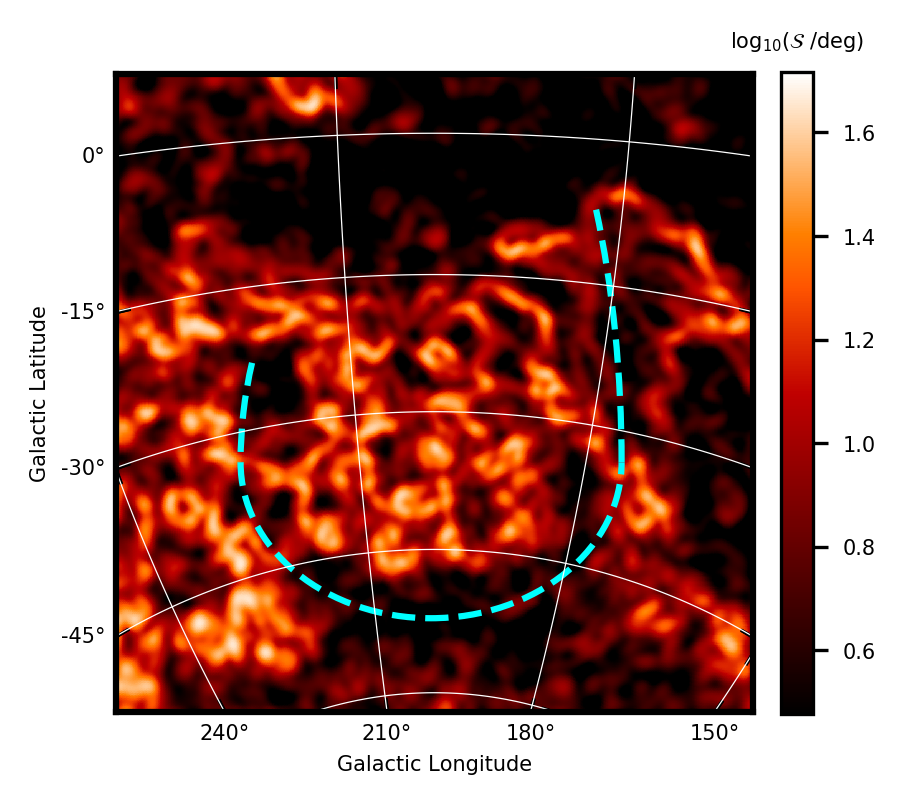}}
\vspace{-0.63cm}
\centerline{\includegraphics[height=0.3\textheight]{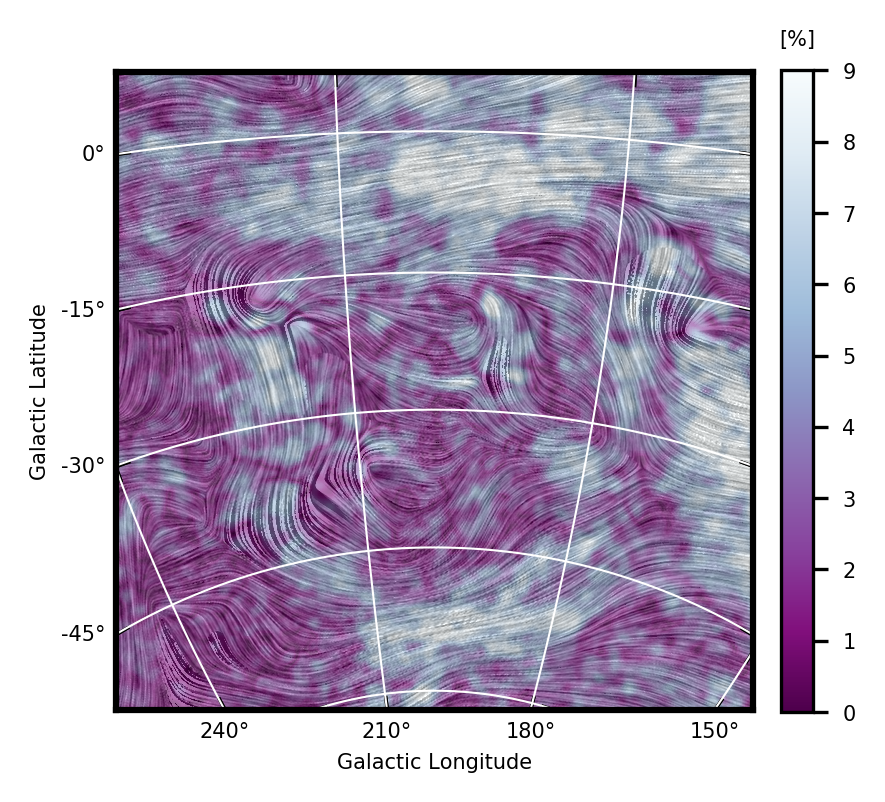}}
\vspace{-0.2cm}
\caption{
Polarization angular dispersion function, $S_{2}(\delta=30\arcm)$ (top) and polarization fraction (bottom), toward the Orion-Eridanus superbubble. 
}\label{fig:smap}
\end{figure}

We quantify the dispersion of the orientation of $\vec{B}_{\perp}$ using the polarization angle dispersion function \citep{hildebrand2009,planck2014-XIX},
\begin{equation}\label{eq:StructureFunction}
\mathcal{S}_{2}(\bm{x}, \bm{\delta}) =  \left( \frac{1}{N} \sum_{i=1}^N \left(\Delta \psi_{xi} \right)^2 \right)^{1/2},
\end{equation}
where $\Delta \psi_{xi} = \psi(\bm{x}) - \psi(\bm{x} + \bm{\delta_i})$ is the difference in polarization angle ($\psi$, see Eq.~\ref{eq:polangle}) between a given position $\bm{x}$ and a position offset by a displacement $\bm{\delta}$.
The \juan{top} panel of Fig.~\ref{fig:smap} presents $\mathcal{S}_{2}(\bm{x}, \delta$\,=\,$30\arcmin)$, \juan{the same $\delta$ value used in \cite{planck2014-XIX}, although our conclusions do not depend on this particular selection (see Appendix~\ref{app:Planck})}. 
\juan{This quantity shows a clear decrease in the polarization angle dispersion along the edge of the superbubble, particularly in the eastern and southern edges, relative to the surroundings}.
\juan{This low $\mathcal{S}_{2}$ values are further accompanied by relatively high polarization fractions ($p$, Eq.~\ref{eq:polangle}), as illustrated in the bottom panel of Fig.~\ref{fig:smap}}.
\juan{Both these observations are expected from the large-scale organization of the magnetic field along the wall of the superbubble, following its expansion} \citep{ferriere1991,planck2015-XXXIV,soler2017b}.

Figures \ref{fig:Bfield} and \ref{fig:smap} show that both Arc~B and Arc~C \juan{are inside and run parallel to the low-$\mathcal{S}_{2}$ and high-$p$ outline} .
\juan{This observation} decisively rules out the skinny-bubble shape proposed by \citet{Pon14komp}, who suggested that Arc\,C is a filament outside of the bubble that is merely ionized by photons penetrating the bubble wall.
The superbubble wall also seems to extend beyond Barnard's Loop, \juan{as suggested by the $\mathcal{S}_{2}$ map}, confirming the suggestion of \cite{Ochsendorf15} that Barnard's Loop is a separate shell nested within the larger Orion-Eridanus superbubble.
\juan{Moreover, since Barnard's Loop is expanding within the cavity evacuated by the superbubble, there was not a lot of neutral matter nor magnetic field lines for it to sweep up}, such that it is unsurprising that it does not show up prominently in Fig.~\ref{fig:smap}.
Unfortunately, due to confusion along the Galactic \juan{plane}, the northern end of the superbubble cannot be definitively located using the polarization data, such that it is unclear whether the bubble extends up to a latitude of $5$\deg\ \citep{Robitaille17Scaife}, instead of only going to $-5$\deg\ \citep{Pon16Ochsendorf}. 

\juan{The $\vec{B}_{\perp}$ orientation shown in Fig.~\ref{fig:Bfield} seems to follow the orientation of Arc~A, although this structure does not appear prominently in Fig.~\ref{fig:smap}.
Distance estimates determined using optical photometry of stars from PanSTARRS-1 \citep{Schlafly14} conclusively place Arc~A within 200\,pc of the Sun and they find no evidence of significant reddening beyond 500\,pc toward this position (18 in their table~2). 
As such, we consider the non-detection of Arc~A in $\mathcal{S}_{2}$ to be due to its the projection and confusion, rather than it being a background feature behind the superbubble \citep{Boumis01, Welsh05}}.
 
In the $\mathcal{S}_{2}$ map, the superbubble appears relatively circular. 
\juan{Such a round superbubble is more consistent with the evolution of a superbubble in an exponential-density-profile atmosphere than the highly elongated models of \cite{Pon14komp}}. 
Therefore, additional processes, such as turbulent shaping and \juan{shear from Galactic differential rotation}, are not required to explain the \juan{observed} morphology, although we do not rule out the possibility that such processes have affected the Orion-Eridanus superbubble.
One inescapable fact that strikes the eye from these polarization measurements is that the Galactic magnetic field and the expanding Orion-Eridanus superbubble have clearly interacted and influenced one another. 

\section{Magnetic field strengths}\label{sec:strength}

\begin{table*}
\centering 
\begin{threeparttable}
\caption{Estimates of the magnetic field strength}\label{table:DCFvalues} 
\begin{tabular}{c c c c c c c c}          
\hline\hline                        
($l$, $b$) & $\sigma_{v}$$^{\rm a}$ & $n_{\rm H}$ & $\varsigma_{\psi}$ & $b(0)$ & $B^\textsc{DCF}_{\perp}$ & $B_{\parallel}$$^{\rm b}$ & $v_\textsc{A}$$^{\rm c,d}$ \\    
                & [km\,s$^{-1}$]    &  [cm$^{-3}$]            & [deg]                                       & [deg]             & [$\mu$G]                       & [$\mu$G]                         & [km\,s$^{-1}$] \\
\hline                                   
(191\pdeg5, $-$50\pdeg5) & 2.3$\pm$0.1 &  70$\pm$14 & \phantom{0}6.9$\pm$0.5 & \phantom{0}6.2$\pm$0.5 & 87$\pm$9 & \phantom{0}$-6.6\pm0.6$ & 19$\pm$4 \\  
(195\pdeg4, $-$50\pdeg2) & 2.0$\pm$0.1 &  35$\pm$7\phantom{0} &\phantom{0}6.7$\pm$0.5 & \phantom{0}7.1$\pm$0.5 & 55$\pm$6  &                   $-11.8\pm1.3$ & 17$\pm$4 \\
(199\pdeg0, $-$50\pdeg2) & 2.2$\pm$0.1 &  50$\pm$10  &          19.2$\pm$0.5 & 20.3$\pm$0.5                   &         25$\pm$3  & \phantom{0}$-9.5\pm0.7$ & \phantom{0}7$\pm$2 \\
\hline                                             
\end{tabular}
\vspace{0.1cm}
\begin{tablenotes}
	\item[a] Estimated from the Gaussian fit to the line profiles presented in Appendix~\ref{app:HIspectra}.
	\item[b] As reported in table 3 of \cite{Heiles89}.
	\item[c] Alfv\'{e}n velocity $v_\textsc{A}$\,$=$\,$B^\textsc{tot}/(4\pi\rho)^{1/2}$, where $B^\textsc{tot}$\,$\equiv$\,$[(B^\textsc{DCF}_{\perp})^{2}+(B_{\parallel})^{2}]^{1/2}$. 
	\item[d] We assume a mean molecular weight $\mu$\,$=$\,$1.4$.
\end{tablenotes}
\end{threeparttable}
\end{table*}

Quantifying the magnetic properties of the whole Orion-Eridanus superbubble is beyond the scope of this letter.
We present an estimate of the magnetic field strength in the region where \juan{polarization observations were previously not available and the line-of-sight confusion is less severe}, that is, $b$\,$<$\,$-30$\deg.
Particularly, we focus on three positions located at the interface of the H$\alpha$ and H{\sc i} shells in the southern edge of the superbubble, where prior measurements of $B_{\parallel}$, based upon the H{\sc i} Zeeman effect in the 21-cm emission line \citep{Heiles89}, are available.
Toward these three positions, illustrated in Fig.~\ref{fig:Bfield}, we complement the observations of $B_{\parallel}$ with the estimates of ${B}_{\perp}$ obtained with the Davis-Chandrasekhar-Fermi method \citep[hereafter the DCF,][]{Davis51,Chandrasekhar53}.

The DCF method provides an estimate of $B_{\perp}$ in the \juan{interstellar medium}.
Under the assumptions that the magnetic field is frozen into the gas and that the dispersion of the local magnetic-field orientation angle is due to transverse and incompressible Alfv\'{e}n waves, 
\begin{equation}\label{eq:bdcf}
B^\textsc{DCF}_{\perp}=\sqrt{4\pi\rho}\frac{\sigma_{v}}{\varsigma_{\psi}},
\end{equation}
where 
$\sigma_{v}$ is the velocity dispersion of the gas, 
$\rho$ is the gas mass density, 
and $\varsigma_{\psi}$\footnote{We use this notation to avoid confusion with $\sigma_{\psi}$, which is the uncertainty in the polarization orientation angle, $\psi$.} is the angular dispersion of the local magnetic-field orientations.
Recent studies of the DCF method based on simulations of magnetohydrodynamical (MHD) turbulence in molecular clouds \citep{Ostriker2001} provide a correction factor of 0.5 to Eq.~\ref{eq:bdcf}. 
We choose to omit this factor due to the dissimilar physical conditions in the considered regions and the MHD simulations employed to estimate this correction factor. 

We estimate $B^\textsc{DCF}_{\perp}$ in 3\deg-diameter \juan{circles centred at} the positions listed in Table~\ref{table:DCFvalues}.
\juan{The angular resolution of the polarization observations (1\deg) is slightly lower than that of the Zeeman observations (0\pdeg6).
As such, we do not probe exactly the same regions in $B^\textsc{DCF}_{\perp}$ as in $B_{\parallel}$. 
Nevertheless, the three Zeeman splitting values are consistent, that is, they reflect $B_{\parallel}$ pointing towards the observer without a reversal. 
Therefore, to first order, we assume that the $B_{\parallel}$ values are representative of the larger region sampled by our analysis.} 

We estimate $\sigma_{v}$ using the H{\sc i} spectra averaged in the 3\deg-diameter \juan{circles} and the spectral decomposition procedure described in Appendix~\ref{app:HIspectra}.
For all three cases, the spectra present narrow components and wide wings associated with an additional underlying, broad, spectral line.
We assume that the H{\sc i} emission comes from a mixture of two stable gas phases \citep{Wolfire03}, the cold neutral medium (CNM) and the warm neutral medium (WNM), which produce the narrow and broad components, respectively. 
Recent studies of dust polarized emission and H{\sc i} data show a strong correlation between magnetic-field and CNM structures, in contrast with the poor correlation with the WNM \citep{Clark2014, planck2014-XXXII}.
Motivated by these results, we choose to only use the CNM component to estimate the gas properties involved in the calculation of $B^\textsc{DCF}_{\perp}$, which we present in Table~\ref{table:DCFvalues}.
If instead of using the CNM component we considered the WNM component, the magnetic-field strengths in Table~\ref{table:DCFvalues} would be larger given its larger H{\sc i} velocity dispersion. 
Thus, our estimates of $B_{\perp}$ constitute lower limits to the true values. 

The volume density, $n_{\rm H}$, is derived from the H{\sc i} column density assuming that the line-of-sight depth, $\Delta$, of the CNM component is of a few parsec. 
A lower limit for $\Delta$ (roughly 1 pc) is set by the negligible molecular gas fraction toward the three positions \citep{planck2013-p03a}.
An upper limit for $\Delta$ is set by the typical densities of CNM gas, $n_{\rm H}$\,$\approx$\,30\,cm$^{-3}$.
For the sake of simplicity, we use $\Delta$\,$\approx$\,5\,$\pm$1\,pc in the three regions. 

The angular dispersion of the local magnetic field orientation, $\varsigma_{\psi}$, is estimated using two methods. 
First, directly computing the angular dispersion from the Stokes $Q$ and $U$ using equations D.5 and D.11 in \cite{planck2015-XXXV}, and second, using the structure function of polarization angles as described in Appendix~\ref{app:houde2009}.
Central coordinates, velocity dispersions, volume densities, angular dispersions of $\vec{B}_{\perp}$, strengths of $\vec{B}_{\perp}$, and Alfv\'{e}n velocities of the three positions are given in Table~\ref{table:DCFvalues}.

\section{Discussion and conclusions}\label{sec:discussion}

The magnetic field strengths within the shell, reported in Table~\ref{table:DCFvalues}, tend to be a factor of a few larger than the average field strength of CNM clouds in the Solar neighborhood \citep[$\sim$\,6\,$\mu$G,][]{Heiles2005}, \juan{as} expected for magnetic fields compressed by expanding \juan{bubbles}. 
The plane-of-the-sky magnetic field, $\vec{B}_{\perp}$, shows obvious signs of interaction with the superbubble. 
The maps of the polarization angular dispersion and polarization fraction clearly highlight the location of the outer wall of the superbubble. 
This outer wall traces out a relatively spherical superbubble, consistent with the expansion of the superbubble in the exponential density profile of the Galactic disk. 
These data rule out the skinny bubble models of \citet{Pon14komp} and are more consistent with the models put forth by \citet{Ochsendorf15} and \cite{Pon16Ochsendorf}. 

The values of $\vec{B}_{\perp}$ and $\vec{B}_{\parallel}$ indicate that the magnetic field vector has a large component in the plane of the sky, as expected from the relatively large polarization fractions observed toward these regions. 
The Alfv\'{e}n velocities in the shell, inferred from the combination of $\vec{B}_{\perp}$ and $\vec{B}_{\parallel}$, and listed in Table~\ref{table:DCFvalues}, are below the shell expansion velocity reported in \citet{Reynolds79}, approximately 23\,km/s, \juan{further} suggesting that the magnetic field was swept up and organized by the expanding \juan{bubble}. 
This is also illustrated by Fig.~\ref{fig:Bfield} and Fig.~\ref{fig:smap}, where $\vec{B}_{\perp}$ lies parallel to the shell surface over a large portion of its projected extent. 
Although it is expected that the initial magnetic field in the region hinders the expansion of the superbubble perpendicular to its mean direction \citep{Tomisaka98}, \juan{the circular shape of the superbubble and} the data in hand indicate that the magnetic field is not dynamically important for the expansion of the superbubble itself.
This is in agreement with results of recent numerical MHD simulations that show that moderate initial magnetic fields do not modify the total amount of momentum injected by SNe explosions into the ISM \citep{kim2015,iffrig2017}.

However, the now-compressed magnetic field may play a significant role in the continuing evolution of the \juan{shell structure}, both generating surfaces more stable to dynamical instabilities and suppressing the formation of dense gas. 
The enhanced magnetic field provides additional support, reducing the dense gas fraction \citep{walch2015,ntormousi2017} and potentially shaping the \juan{photoionization region} produced by the generation of stars that follow the SN explosion \citep{Basu99,pellegrini2007}. 
The investigation of these effects motivates the pursuit of additional observations of Zeeman splitting and \juan{dust polarized emission at higher angular resolution} toward the Orion-Eridanus and other superbubbles.

\begin{acknowledgements}
JDS acknowledges the support from the European Research Council (ERC) under the Horizon 2020 Framework Program via the Consolidator Grant CSF-648505. 
AB acknowledges the support from the ERC Advanced Grant ORISTARS under the European Union's Seventh Framework Programme (Grant 291294). 
AP was partly funded by a Canadian Institute for Theoretical Astrophysics (CITA) National Fellowship. 
We thank the referee M.~Alves for her constructive comments that have led to improvements in this presentation of our results.
Thanks to M. Houde and S. Basu for their comments on the early versions of this work.
We also thank the following people who helped with their encouragement and conversation: P.~G.~Martin, H.~Beuther, C.~Heiles, T.~Robishaw, M.-A. Miville-Desch\^{e}nes, and A.~Roy.
Special thanks go to Thomas M\"{u}ller at the Haus der Astronomie for his assistance in the magnetic field visualization.
\end{acknowledgements}
 
\bibliographystyle{aa}
\bibliography{planckbubblevPostAandAReport2.bbl}

\appendix 

\section{\Planck\ polarization maps}\label{app:Planck}

\begin{figure}[ht!]
\vspace{-0.4cm}
\centerline{\includegraphics[width=0.42\textwidth]{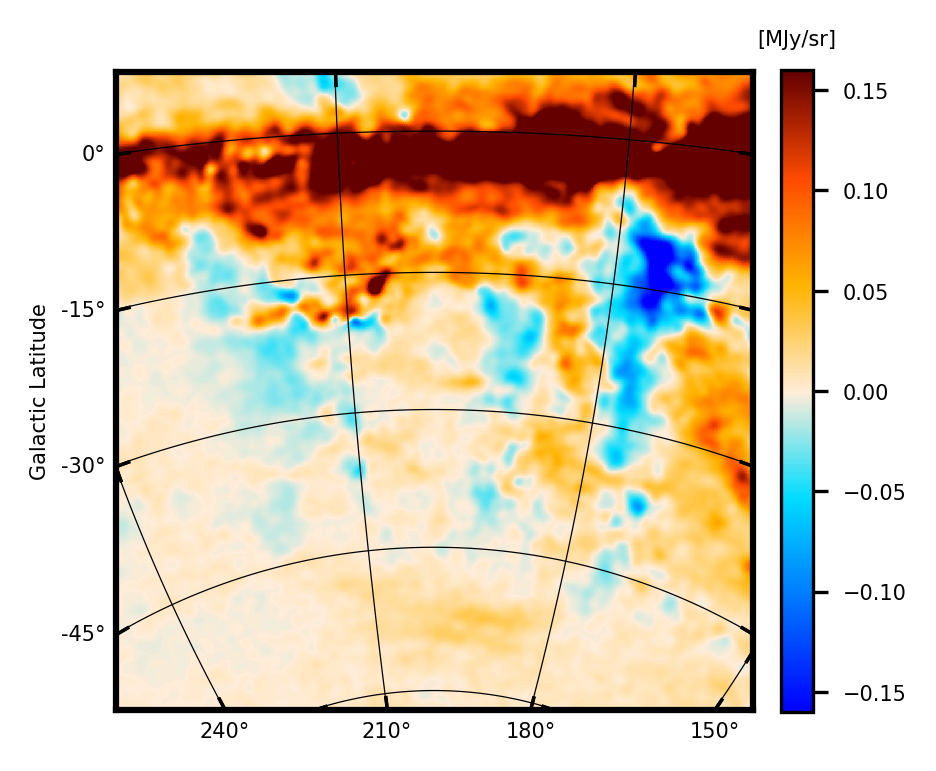}}
\vspace{-0.47cm}
\centerline{\includegraphics[width=0.42\textwidth]{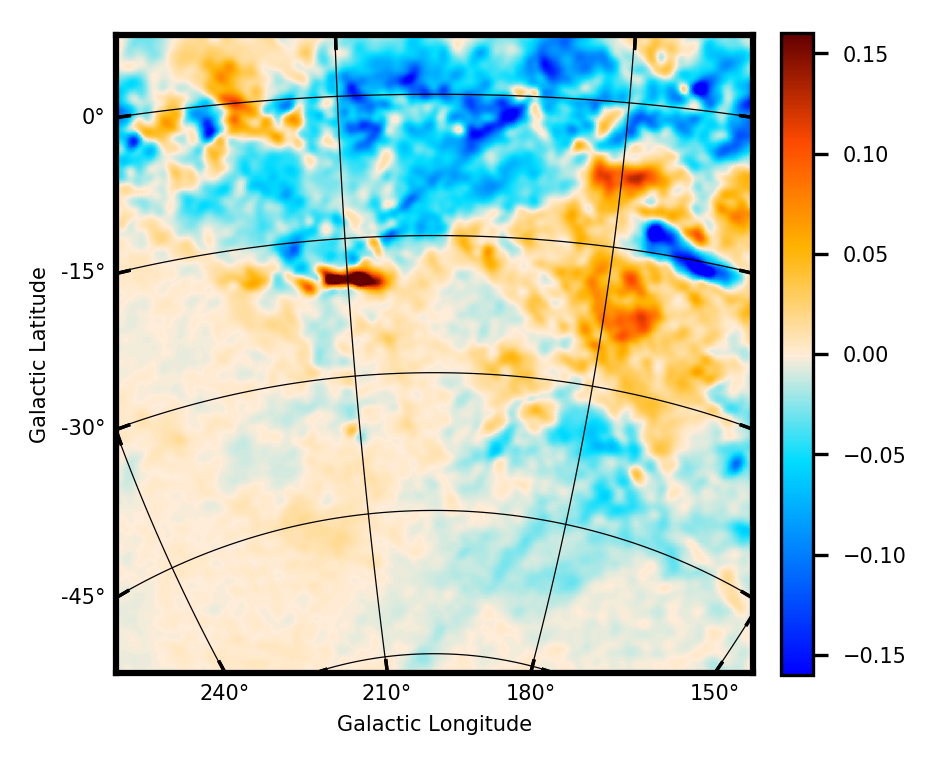}}
\vspace{-0.1cm}
\caption{\Planck\ 353-GHz Stokes $Q$ (top) and $U$ (bottom) maps toward the Orion-Eridanus superbubble at 1\deg\ FWHM resolution.
}\label{fig:QandUmap}
\end{figure}

\subsection{Stokes $I$, $Q$ and $U$ maps}

We use the publicly available 353-GHz Stokes, $I$, $Q$ and $U$ maps and the associated noise maps made with the five independent consecutive sky surveys of the \Planck\ cryogenic mission. 
We refer to publications by the \Planck\ Collaboration for details \juan{on the data processing}, including mapmaking, photometric calibration, and photometric uncertainties \citep[][and references therein]{planck2014-a01}. 
The $Q$ and $U$ maps are initially constructed with an effective beamsize of 4\parcm8. 
The three maps are in {\tt HEALPix} format \citep{gorski2005} with a pixelization at $N_{\rm side}$\,$=$\,$2048$, which corresponds to an effective pixel size of 1\parcm7. 
To guarantee a signal-to-noise ratio \juan{greater than} 3 in the extended polarized emission observations over the whole region, we smooth the \planck\ 353-GHz maps to 1\deg\ resolution using a Gaussian approximation to the \Planck\ beam.
The smoothing is produced with the {\tt ismoothing} routine of {\tt HEALPix}, which decomposes the $Q$ and $U$ maps into $E$ and $B$ maps, applies Gaussian smoothing in harmonic space, and transforms the smoothed $E$ and $B$ back into $Q$ and $U$ maps. 

We compute the polarization angles and fractions used for the analysis using
\begin{equation}\label{eq:polangle}
\psi = \frac{1}{2}\arctan(-U,Q) {\,\,\,\rm and \,\,\,} p = \frac{\sqrt{Q^2+U^2}}{I}, 
\end{equation}
respectively.
The minus sign is needed to change the {\tt HEALPix} format maps 
into the International Astronomical Union (IAU) convention for $\psi$, measured from the local direction to the north Galactic pole with increasing positive values toward the east.

\subsection{Dispersion function maps}

\begin{figure}[ht!]
\vspace{-0.2cm}
\centerline{\includegraphics[width=0.42\textwidth]{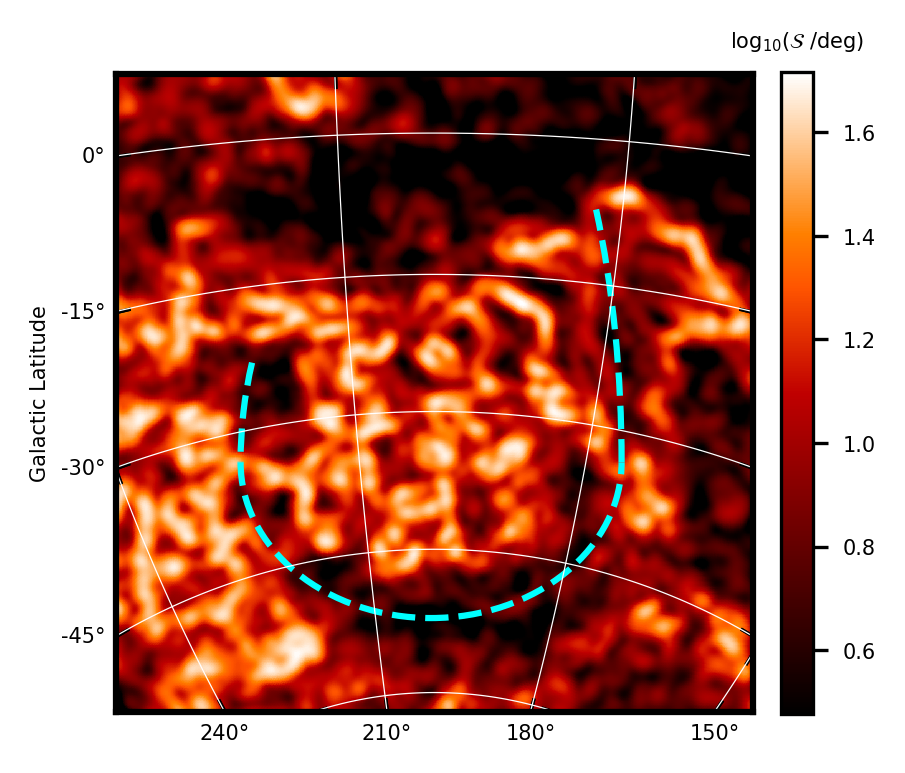}}
\vspace{-0.47cm}
\centerline{\includegraphics[width=0.42\textwidth]{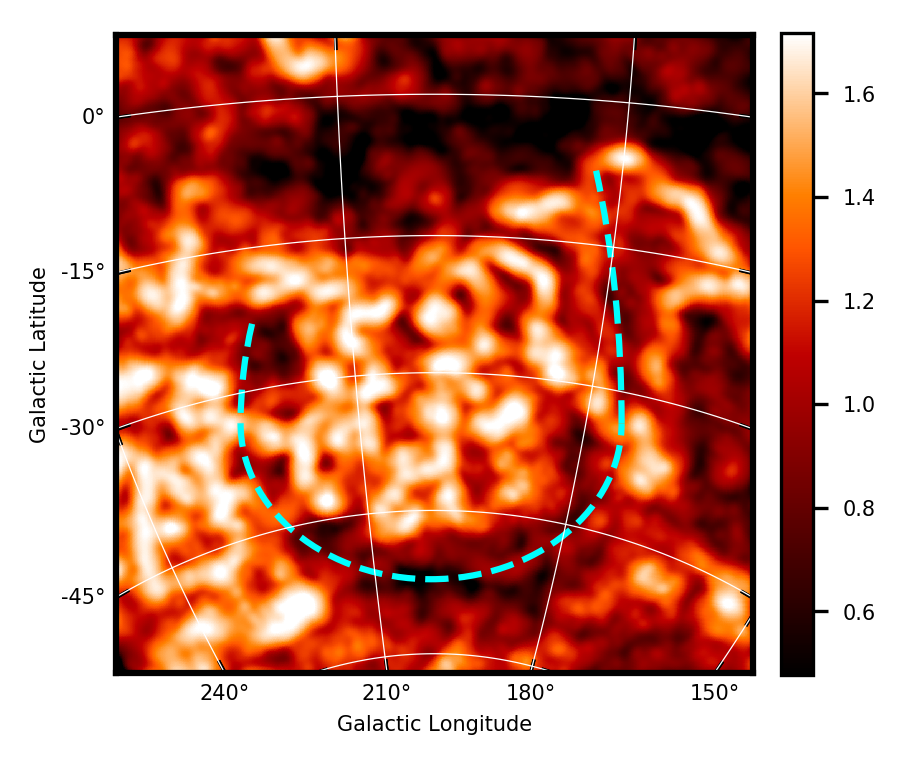}}
\vspace{-0.2cm}
\caption{\juan{Polarization angular dispersion function, $S_{2}(\bm{x}, \bm{\delta})$, Eq.~\ref{eq:StructureFunction}, towards the Orion-Eridanus superbubble. 
The top and bottom panels correspond to $S_{2}(\delta=60\arcm)$ and $S_{2}(\delta=90\arcm)$, respectively}.
}\label{fig:smapMultiDelta}
\end{figure}

\juan{In the main body of this letter, we considered the particular case $\mathcal{S}_{2}(\bm{x}, \delta$\,=\,$30\arcmin)$.
Here, for the sake of comparison, we present in Fig.~\ref{fig:smapMultiDelta} the maps of $\mathcal{S}_{2}(\bm{x}, \bm{\delta})$ computed at displacements $\delta=60$\arcm\ and $90$\arcm.
Both panels show that the low-$\mathcal{S}_{2}$ region identified in Fig.~\ref{fig:smap} and associated with the wall of the expanding superbubble is also visible in the maps obtained with larger values of $\delta$, thus confirming that it is not the product of the correlation introduced by the resolution of the observations.}

\subsection{Gradient of polarization maps}

\juan{The computation of extended $\mathcal{S}_{2}(\bm{x}, \bm{\delta})$ maps is costly and time consuming.
An alternative approach is computing the quantity}
\begin{equation}\label{eq:gradPoverP}
\frac{|\nabla P|}{P}=\frac{1}{\sqrt{Q^2 + U^2}}\left[\left(\frac{\partial Q}{\partial x}\right)^{2}+\left(\frac{\partial Q}{\partial y}\right)^{2}+\left(\frac{\partial U}{\partial x}\right)^{2}+\left(\frac{\partial U}{\partial y}\right)^{2}\right]^{1/2},
\end{equation}
\juan{which was introduced in \cite{gaensler2011} and traces the changes in the orientation of $B_{\perp}$}.

\begin{figure}[ht!]
\vspace{-0.3cm}
\centerline{\includegraphics[width=0.42\textwidth]{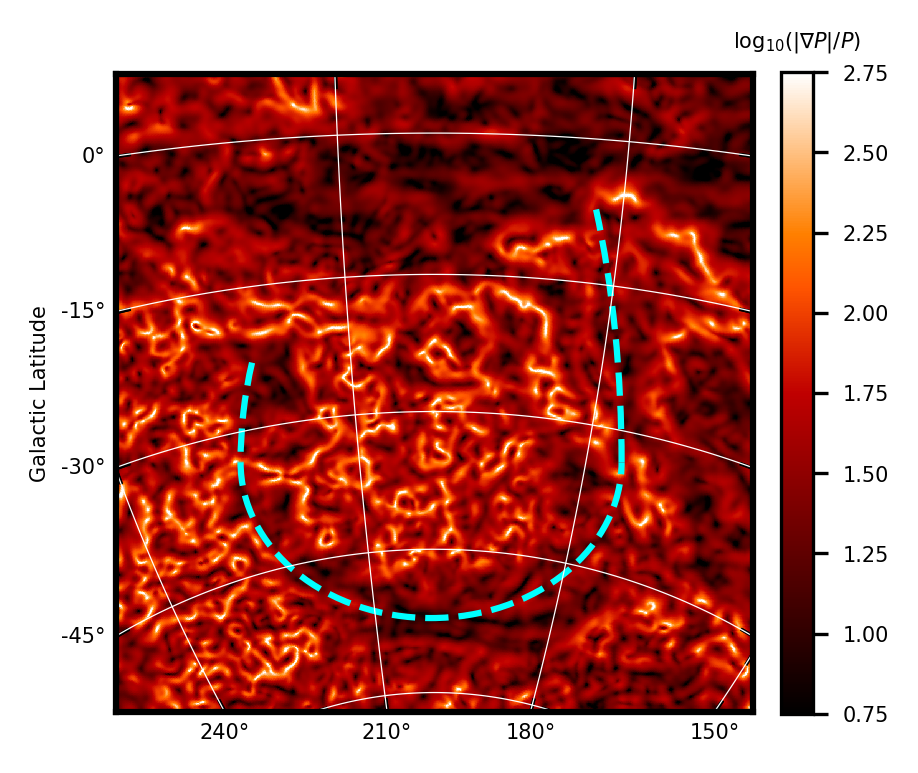}}
\vspace{-0.46cm}
\centerline{\includegraphics[width=0.42\textwidth]{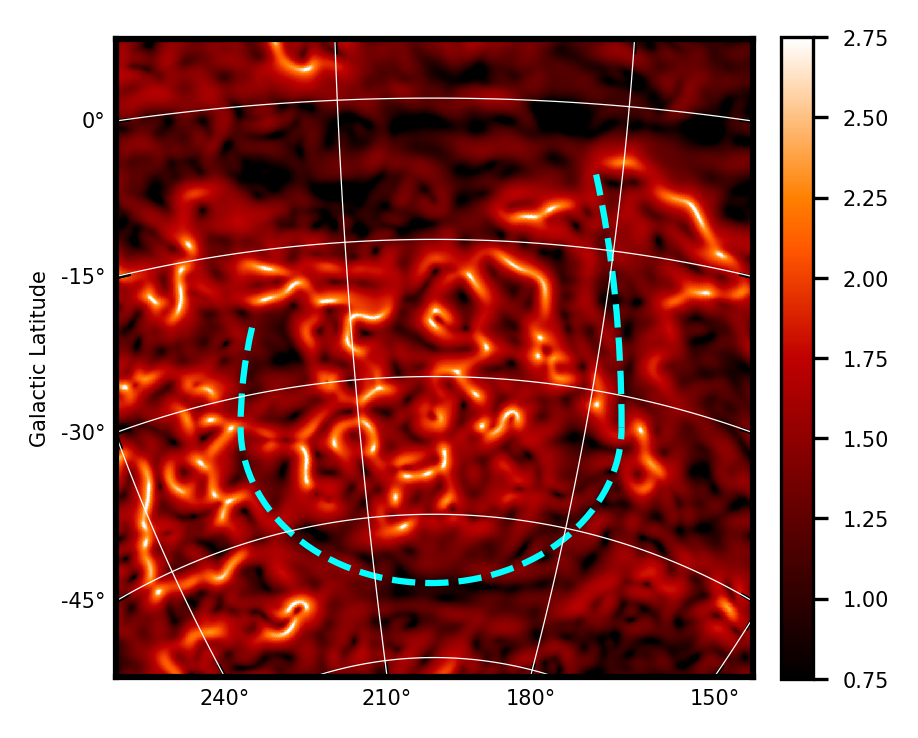}}
\vspace{-0.455cm}
\centerline{\includegraphics[width=0.42\textwidth]{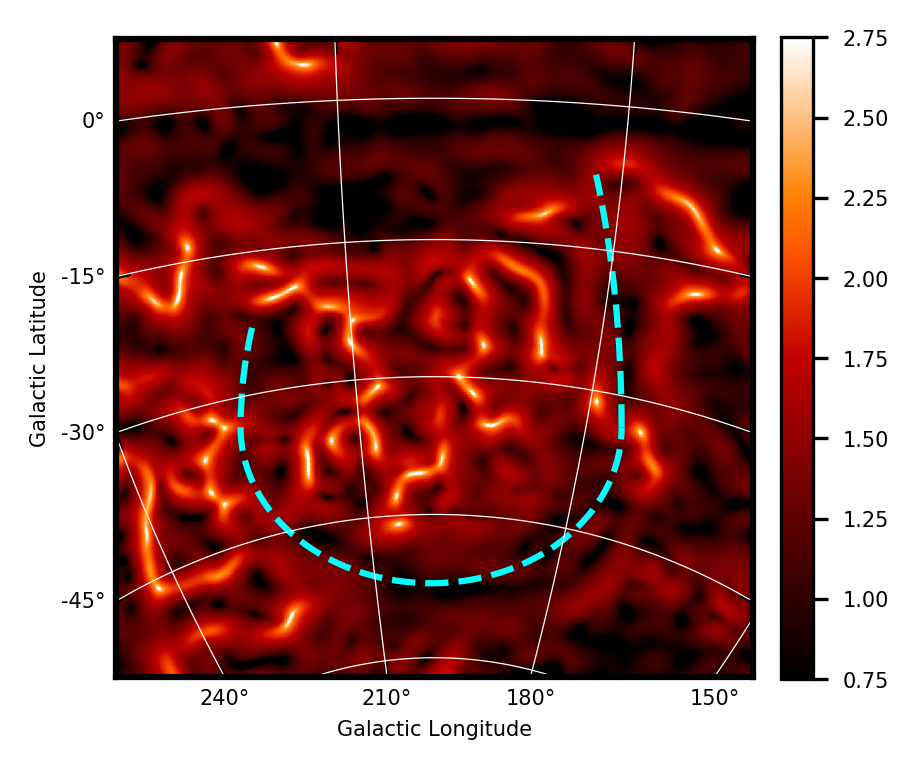}}
\vspace{-0.1cm}
\caption{Maps of $|\nabla P|/P$, Eq.~\ref{eq:gradPoverP}, toward the Orion-Eridanus superbubble. The spatial gradients are calculated using derivative kernels of diameters 60\arcm\ (top), 120\arcm\ (middle), and 180\arcm\ (bottom).}\label{fig:GradPoverP}
\end{figure}

\juan{Figure~\ref{fig:GradPoverP} shows maps of $|\nabla P|/P$ where the spatial derivatives of Stokes $Q$ and $U$ are calculated using derivative kernels with diameters 60\arcm, 120\arcm, and 180\arcm, as introduced in \cite{soler2013}.
In practice, this means that we smooth the data with a two-dimensional Gaussian with FWHM equal to the corresponding diameter and then calculate the gradient using next-neighbour differences. 
This operation is equivalent to calculate the derivative using differences of points inside a vicinity with the same diameter.
The diameter of the derivative kernel is roughly comparable to the displacement $\delta$ in $\mathcal{S}_{2}(\bm{x}, \bm{\delta})$.
The $|\nabla P|/P$ maps show that the low-polarization-angle-dispersion region identified in Fig.~\ref{fig:smap} is also visible in $|\nabla P|/P$ at scales around and above the angular resolution of the observations.}

\section{Gaussian decomposition of H{\sc i} spectra}\label{app:HIspectra}

\begin{figure}
\centering
\vspace{-0.1cm}
\includegraphics[width=0.48\textwidth]{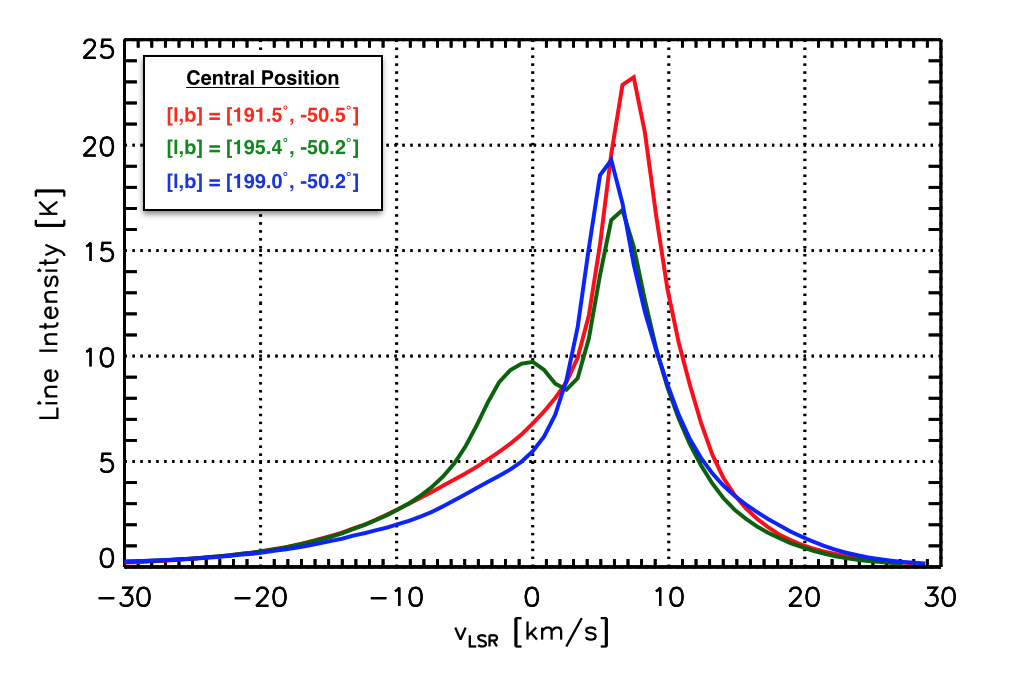}
\vspace{-0.4cm}
\caption{Average spectra in 3\deg-diameter vicinities centered on the 21-cm emission-line Zeeman splitting observations presented in \cite{Heiles89}.}\label{fig:DCFregions}
\end{figure}

\begin{figure}[ht!]
\centerline{\includegraphics[width=0.45\textwidth]{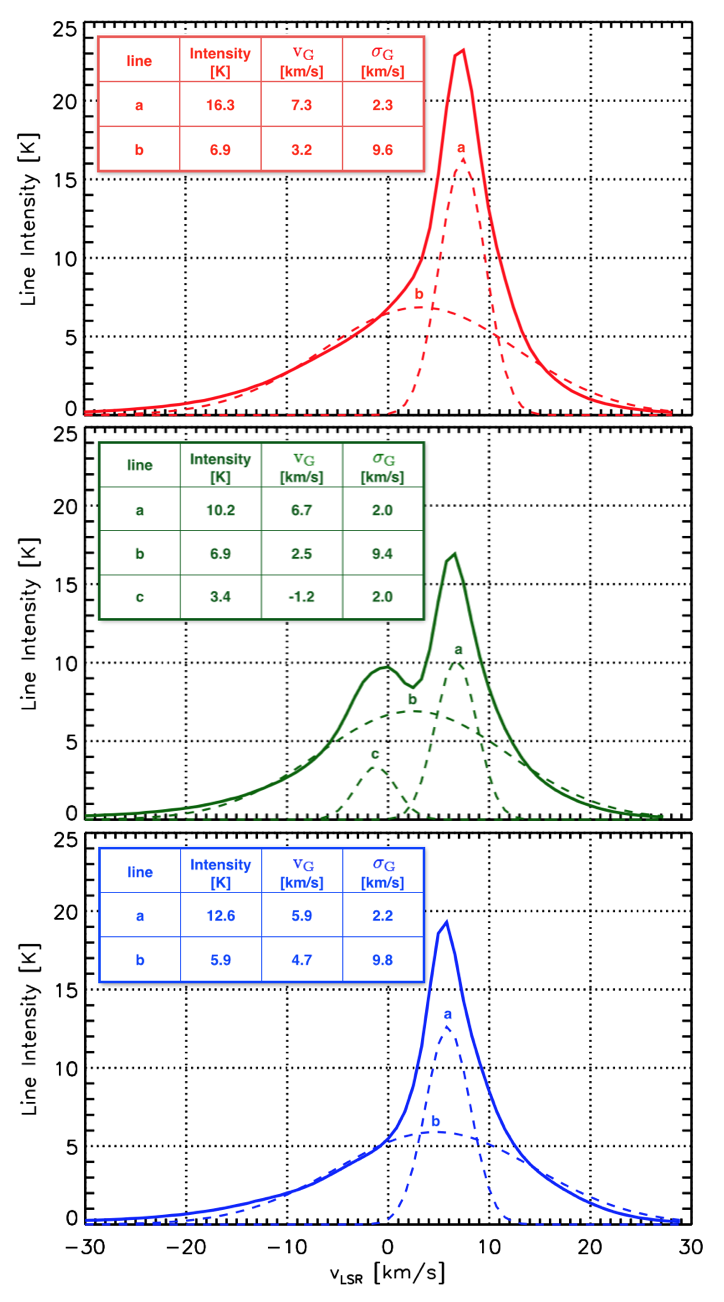}}
\vspace{-0.2cm}
\caption{
Gaussian decomposition of the H{\sc i} spectra presented in Fig.~\ref{fig:DCFregions} with colours representing the three central positions. 
The solid lines correspond to full average spectra while dashed lines to the decomposed Gaussians. 
The parameters of each Gaussian component: intensity, centroid velocity ($v_{\rm G}$), and standard deviation ($\sigma_{\rm G}$), are listed in each panel.}
\label{fig:GaussHI}
\end{figure}

In order to account for the multiphase nature of the H{\sc i} emission \citep{Wolfire03} in our implementation of the DCF method, we apply a simple Gaussian decomposition of the spectra.
Assuming a bistable atomic gas, we separate the emission from the cold neutral medium (CNM) and the warm neutral medium (WNM). 
For the three spectra in Fig.~\ref{fig:GaussHI}, we show the decomposed Gaussians with the key parameters in the corresponding tables. 
In all cases, we find narrow spectral lines with $\sigma_{\rm G}$\,$\sim$\,2\,km/s, typical of CNM emission, on top of broad WNM line with $\sigma_{\rm G}$\,$>$\,9\,km/s. 
In this letter, we focus on the narrow CNM component (line \emph{a} in Fig.~\ref{fig:GaussHI}) at positive velocities for which \citet{Heiles89} estimated the strengths of $B_{\parallel}$ from Zeeman measurements. 

We estimate the CNM column density for the three different regions using the relation
\begin{equation}\label{eq:nh} 
\frac{N_{\rm H}}{\rm cm^{-2}}\approx 1.82\times10^{18} \int\frac{T_{\rm b}(v)}{\rm K}\frac{dv}{\rm km/s},
\end{equation}
where $T_{\rm b}$ is the observed 21-cm line brightness temperature (line intensity in the figures) at radial velocity $v$ with respect to the local standard of rest \citep{kulkarni1987}.

\section{Dispersion function of polarization angles}\label{app:houde2009}

\begin{figure}
\centering
\vspace{-0.1cm}
\centerline{\includegraphics[width=0.47\textwidth]{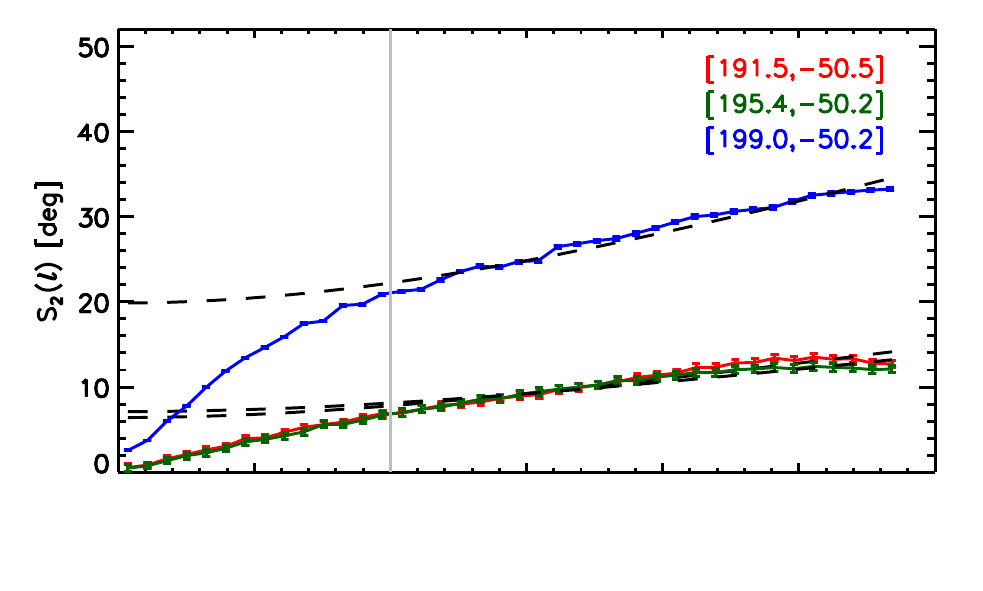}}
\vspace{-1.0cm}
\centerline{\includegraphics[width=0.47\textwidth]{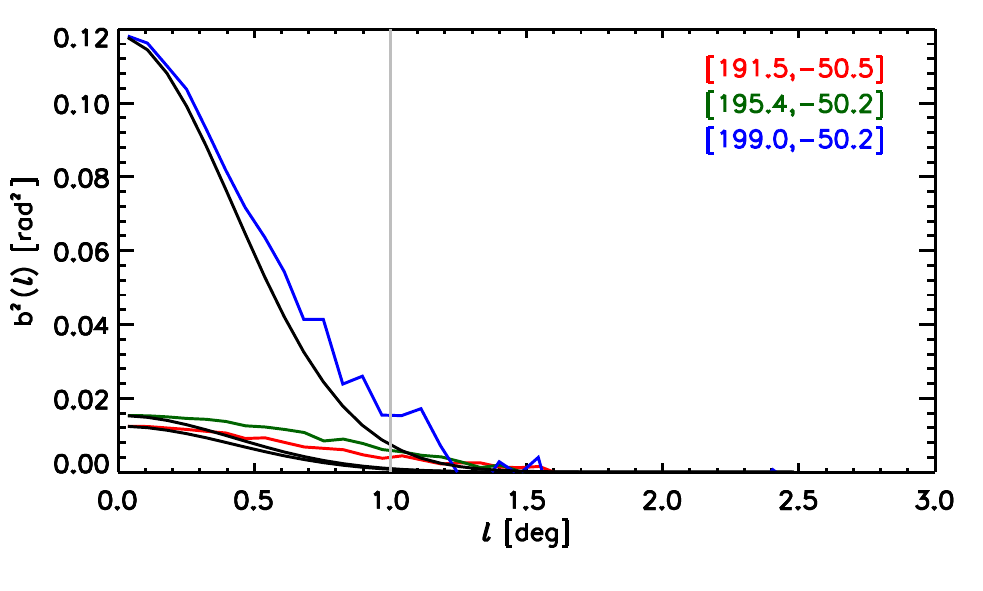}}
\vspace{-0.4cm}
\caption{Structure function of the polarization angles, $S_{2}(\ell)$, and normalized autocorrelation function of the plane-of-the-sky component of the magnetic field, ${b}^{2}(\ell)$ \citep{houde2009}, in the 3\deg-diameter vicinities centered on the 21-cm emission-line Zeeman splitting observations presented in \cite{Heiles89}.
In the top panel, the dashed lines correspond to the fit to the values of $S_{2}(\ell)$ in the range $\ell > 1$\deg, which is the effective resolution of the \planck\ observations, represented by the vertical solid line.
In the bottom panel, the solid black lines correspond to the expected correlation produced by the effective resolution of the observations.
}
\label{fig:StrucFunc}
\end{figure}

\juan{The structure function, $S_{2}(\ell)$; which is simply the dispersion function of the polarization angles (Eq.~\ref{eq:StructureFunction}) averaged over all the positions $\vec{x}$, such that $S_{2}(\ell)$\,$\equiv$\,$\left<\mathcal{S}_{2}(\bm{x}, \bm{\delta})\right>_{\vec{x}}$; is useful} for the evaluation of the dispersion of polarization angles, $\varsigma_{\psi}$, while avoiding the effect of large-scale non-turbulent perturbations \citep{hildebrand2009}.
The evaluation of $\varsigma_{\psi}$ is made by fitting $S^{2}_{2}(\ell)$ with a second-order polynomial, $S^{2}_{2}(\ell)=b(\ell)+a'_{2}\ell^{2}$, and evaluating the intercept, $b(0)$.
The values of $S_{2}(\ell)$ and the aforementioned fits are presented in the top panel of Fig.~\ref{fig:StrucFunc}. 

In principle, \juan{the DCF magnetic field estimates can be corrected for the effect of line-of-sight integration} if one can reliably estimate of number of independent turbulent cells along the line of sight, $N$ \citep{houde2009}.
Unfortunately, the correlation introduced by the effective \Planck\ beam dominates the values of $b(\ell)$, as illustrated in the bottom panel of Fig.~\ref{fig:StrucFunc}, and it is not possible to determine the characteristic scale of turbulence \citep[see][for a detailed discussion]{MivilleDeschenes2016} necessary to estimate $N$ from these data alone.

\end{document}

%% file: Planck.tex
\def\setsymbol#1#2{\expandafter\def\csname #1\endcsname{#2}}
\def\getsymbol#1{\csname #1\endcsname}

\def\Planck{\textit{Planck}}

\def\all2013resultspapers{\nocite{planck2013-p01, planck2013-p02, planck2013-p02a, planck2013-p02d, planck2013-p02b, planck2013-p03, planck2013-p03c, planck2013-p03f, planck2013-p03d, planck2013-p03e, planck2013-p01a, planck2013-p06, planck2013-p03a, planck2013-pip88, planck2013-p08, planck2013-p11, planck2013-p12, planck2013-p13, planck2013-p14, planck2013-p15, planck2013-p05b, planck2013-p17, planck2013-p09, planck2013-p09a, planck2013-p20, planck2013-p19, planck2013-pipaberration, planck2013-p05, planck2013-p05a, planck2013-pip56, planck2013-p06b, planck2013-p01a}}

\newbox\tablebox    \newdimen\tablewidth
\def\leaderfil{\leaders\hbox to 5pt{\hss.\hss}\hfil}
\def\tablenote#1 #2\par{\begingroup \parindent=0.8em
    \abovedisplayshortskip=0pt\belowdisplayshortskip=0pt
    \noindent
    $$\hss\vbox{\hsize\tablewidth \hangindent=\parindent \hangafter=1 \noindent
    \hbox to \parindent{$^#1$\hss}\strut#2\strut\par}\hss$$
    \endgroup}

\def\L2{\ifmmode L_2\else $L_2$\fi}

\def\DeltaT{\ifmmode \Delta T\else $\Delta T$\fi}
\def\deltat{\ifmmode \Delta t\else $\Delta t$\fi}
\def\fknee{\ifmmode f_{\rm knee}\else $f_{\rm knee}$\fi}
\def\Fmax{\ifmmode F_{\rm max}\else $F_{\rm max}$\fi}
\def\solar{\ifmmode{\rm M}_{\mathord\odot}\else${\rm M}_{\mathord\odot}$\fi}
\def\Msolar{\ifmmode{\rm M}_{\mathord\odot}\else${\rm M}_{\mathord\odot}$\fi}
\def\Lsolar{\ifmmode{\rm L}_{\mathord\odot}\else${\rm L}_{\mathord\odot}$\fi}
\def\inv{\ifmmode^{-1}\else$^{-1}$\fi}
\def\mo{\ifmmode^{-1}\else$^{-1}$\fi}
\def\sup#1{\ifmmode ^{\rm #1}\else $^{\rm #1}$\fi}
\def\expo#1{\ifmmode \times 10^{#1}\else $\times 10^{#1}$\fi}
\def\,{\thinspace}
\def\lsim{\mathrel{\raise .4ex\hbox{\rlap{$<$}\lower 1.2ex\hbox{$\sim$}}}}
\def\gsim{\mathrel{\raise .4ex\hbox{\rlap{$>$}\lower 1.2ex\hbox{$\sim$}}}}

\def\simprop{\mathrel{\raise .4ex\hbox{\rlap{$\propto$}\lower 1.2ex\hbox{$\sim$}}}}
\def\deg{\ifmmode^\circ\else$^\circ$\fi}
\def\pdeg{\ifmmode $\setbox0=\hbox{$^{\circ}$}\rlap{\hskip.11\wd0 .}$^{\circ}
          \else \setbox0=\hbox{$^{\circ}$}\rlap{\hskip.11\wd0 .}$^{\circ}$\fi}
\def\arcs{\ifmmode {^{\scriptstyle\prime\prime}}
          \else $^{\scriptstyle\prime\prime}$\fi}
\def\arcm{\ifmmode {^{\scriptstyle\prime}}
          \else $^{\scriptstyle\prime}$\fi}
\newdimen\sa  \newdimen\sb
\def\parcs{\sa=.07em \sb=.03em
     \ifmmode \hbox{\rlap{.}}^{\scriptstyle\prime\kern -\sb\prime}\hbox{\kern -\sa}
     \else \rlap{.}$^{\scriptstyle\prime\kern -\sb\prime}$\kern -\sa\fi}
\def\parcm{\sa=.08em \sb=.03em
     \ifmmode \hbox{\rlap{.}\kern\sa}^{\scriptstyle\prime}\hbox{\kern-\sb}
     \else \rlap{.}\kern\sa$^{\scriptstyle\prime}$\kern-\sb\fi}
\def\ra[#1 #2 #3.#4]{#1\sup{h}#2\sup{m}#3\sup{s}\llap.#4}
\def\dec[#1 #2 #3.#4]{#1\deg#2\arcm#3\arcs\llap.#4}
\def\deco[#1 #2 #3]{#1\deg#2\arcm#3\arcs}
\def\rra[#1 #2]{#1\sup{h}#2\sup{m}}

\def\dots{\relax\ifmmode \ldots\else $\ldots$\fi}
\def\WHzsr{\ifmmode $W\,Hz\mo\,sr\mo$\else W\,Hz\mo\,sr\mo\fi}
\def\mHz{\ifmmode $\,mHz$\else \,mHz\fi}
\def\GHz{\ifmmode $\,GHz$\else \,GHz\fi}
\def\mKs{\ifmmode $\,mK\,s$^{1/2}\else \,mK\,s$^{1/2}$\fi}
\def\muKs{\ifmmode \,\mu$K\,s$^{1/2}\else \,$\mu$K\,s$^{1/2}$\fi}
\def\muKRJs{\ifmmode \,\mu$K$_{\rm RJ}$\,s$^{1/2}\else \,$\mu$K$_{\rm RJ}$\,s$^{1/2}$\fi}
\def\muKHz{\ifmmode \,\mu$K\,Hz$^{-1/2}\else \,$\mu$K\,Hz$^{-1/2}$\fi}
\def\MJysr{\ifmmode \,$MJy\,sr\mo$\else \,MJy\,sr\mo\fi}
\def\MJysrmK{\ifmmode \,$MJy\,sr\mo$\,mK$_{\rm CMB}\mo\else \,MJy\,sr\mo\,mK$_{\rm CMB}\mo$\fi}
\def\microns{\ifmmode \,\mu$m$\else \,$\mu$m\fi}

\def\muK{\ifmmode \,\mu$K$\else \,$\mu$\hbox{K}\fi}
\def\microK{\ifmmode \,\mu$K$\else \,$\mu$\hbox{K}\fi}
\def\muW{\ifmmode \,\mu$W$\else \,$\mu$\hbox{W}\fi}
\def\kms{\ifmmode $\,km\,s$^{-1}\else \,km\,s$^{-1}$\fi}
\def\kmsMpc{\ifmmode $\,\kms\,Mpc\mo$\else \,\kms\,Mpc\mo\fi}
\providecommand{\sorthelp}[1]{}